\documentclass[11pt]{article}

\usepackage[utf8]{inputenc}
\usepackage[T1]{fontenc}
\usepackage{graphicx}
\usepackage{amsmath, amssymb}
\usepackage{authblk}
\usepackage{geometry}
\usepackage{hyperref}
\usepackage{caption}
\title{Turning a Disposable Bronchoscope into a Dynamic Speckle Imaging Tool: Yes, It Works.}

\author[1]{Aurélien Plyer}
\author[1]{Elise Colin}
\affil[1]{Department of Information Processing, Université Paris Saclay, Palaiseau, France}
\author[2]{Enrique Garcia-Caurel}
\affil[2]{LPICM, Ecole Polytechnique, IPP, Palaiseau, France}

\date{}

\begin{document}

\maketitle

\begin{abstract}
Dynamic speckle imaging, typically used in laser-illuminated surface diagnostics, has proven valuable for assessing biological activity. In this work, we demonstrate its feasibility in an endoscopic context using a disposable bronchoscope. Despite technical limitations and aliasing artifacts, our preliminary results show discernible vascular structures, indicating potential for minimally invasive diagnostic applications. It is important to note that the imaging systems used in this study are designed primarily for clinical robustness and classical imaging, including single-use sterility, ease of handling, and real-time visualization, and not for scientific fidelity of visual data or computational post-processing. As such, they are not inherently suited to dynamic speckle analysis, which requires precise control over temporal acquisition parameters, linear response characteristics of the imaging sensor, and stable illumination conditions, particularly from the coherent laser source. Nevertheless, our results demonstrate that, even within these constraints, dynamic speckle imaging is indeed achievable. This opens the door to further adaptation and optimization of such clinical imaging tools for functional biomedical investigations.
\end{abstract}

\section{Introduction}

Dynamic speckle imaging is an optical technique that exploits the temporal fluctuations of interference patterns—known as speckles—produced when coherent light, typically from a laser source, is scattered by a complex and dynamic surface. When such illumination is applied to any moving sample, such as living biological tissues, the microscopic movements of scatterers within them (e.g., blood flow, cellular activity, tissue motion) induce subtle but measurable variations in the speckle pattern over time \cite{briers2013laser}. These fluctuations carry valuable information about local physiological processes, enabling real-time imaging of activity at or near the surface.

In biomedical optics, dynamic speckle analysis has found applications primarily in in vivo settings where the biological surface is directly accessible. This includes, for example: \begin{itemize} \item •	Superficial imaging of skin for dermatological assessments. \item Visualization of perfused organs maintained \textit{ex vivo} \cite{plyer2023imaging}, or exposed during surgical procedures. \item Functional monitoring of tissues revealed by operative opening, such as brain or organ surfaces during neurosurgery or abdominal surgery. \end{itemize}

These applications share a common feature: the illuminated tissue must be optically accessible in a relatively stable geometric configuration. However, such constraints limit the applicability of dynamic speckle imaging in minimally invasive or internal diagnostic procedures.
To address this limitation, we explore the potential for extending dynamic speckle imaging to endoscopic systems. Endoscopy offers a flexible platform to access internal anatomical structures without surgical exposure, making it an appealing candidate for bringing speckle-based functional imaging to otherwise inaccessible regions. In this preliminary study, we demonstrate the feasibility of acquiring speckle patterns via a commercial, single-use bronchoscope, and we analyze the challenges and opportunities associated with this novel configuration.

\section{Materials and Methods}

\subsection{Endoscopic Imaging Setup}

\begin{figure}
    \centering
    \includegraphics[width=0.5\linewidth]{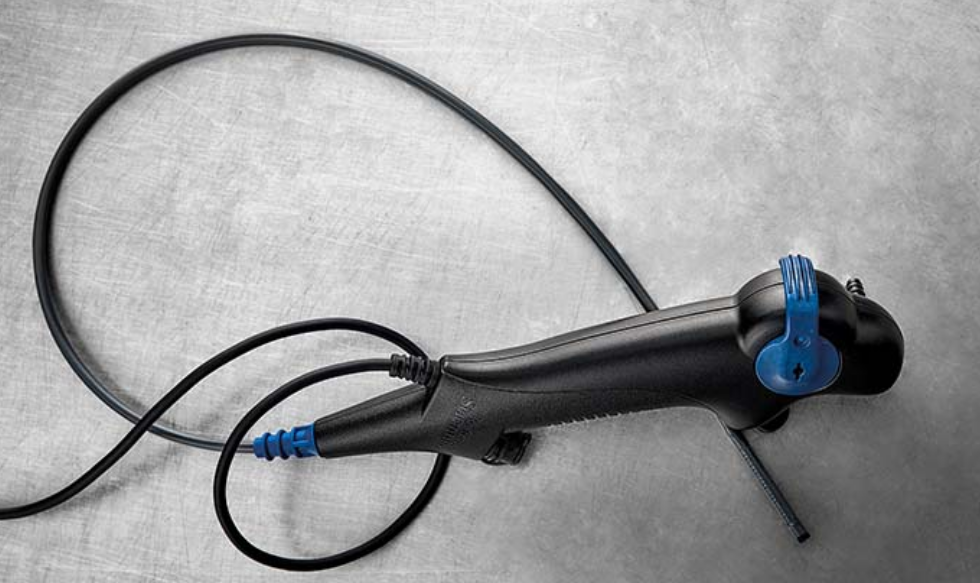}
    \caption{{EXALT\texttrademark~Model B} disposable bronchoscope, developed by Boston Scientific, chosen for our experiment }
    \label{fig:exalt}
\end{figure}

We employed the EXALT Model B disposable bronchoscope, developed by Boston Scientific and released in 2020, as the imaging conduit for dynamic speckle acquisition (Figure \ref{fig:exalt}). This single-use instrument integrates three key functional components at its distal tip: (i) a miniature digital camera, (ii) two white-light LEDs for standard illumination, and (iii) a working channel within the endoscope shaft providing physicians direct access to proven organs. The outer diameter of the “Slim” variant used in this study is approximately 3.8 mm, with an internal channel sufficient to accommodate additional tools or instrumentation.
In our experimental configuration, the two embedded LEDs were turned off, and the default illumination system of the bronchoscope was turned off. Instead, we introduced a standard multimode optical fiber (OM4, core/cladding diameter 50/125 $\mu$m) through the working channel to deliver coherent light directly to the distal end of the endoscope. The light source used was a fiber-coupled 685 nm laser diode operating at 15 mW, initially designed and sold to test optical fiber connections in the telecommunication sector.. Additional tests were performed at higher powers (30 mW and 60 mW), but no significant improvement in image contrast was observed. On the contrary, excessive illumination posed a risk of saturating the red channel of the imaging sensor, thereby degrading speckle contrast. For this reason, the 15 mW setting was maintained for all subsequent acquisitions.
The internal camera of the EXALT bronchoscope is encapsulated within a sealed distal housing and connected to the external display unit via a proprietary digital cable. The camera likely employs a miniaturized CMOS sensor. However, as these components are proprietary to the manufacturer, no detailed public documentation is available regarding the sensor specifications (e.g., pixel size, bit depth, lens characteristics, or signal format). Technical patents related to the EXALT system (e.g., US10258499B2, US10485271B2 \cite{boston_patents}) suggest the use of integrated video compression, most likely H.264 or MPEG, optimized for real-time clinical display.
The bronchoscope is connected to a dedicated Boston Scientific monitor, which reports a maximum display resolution of 2736x1824 pixels. However, this figure likely corresponds to an upscaled display output, not the camera's native resolution. Moreover, the acquisition software and firmware of the display unit are inaccessible to the user, leaving no option for direct retrieval of raw video frames or metadata.
To circumvent these limitations, we implemented a real-time screen capture approach. Frames were recorded directly from the monitor at 60 Hz using screen-grabbing techniques, and subsequently cropped to extract the region of interest containing the live endoscopic image. Although this method introduces limitations regarding compression control and temporal synchronization, it provided a pragmatic solution for time-resolved speckle analysis without access to the internal acquisition pipeline.

\subsection{Speckle Analysis and Signal Considerations}

Several limitations inherent to the imaging pipeline significantly constrained our ability to characterize the speckle signal with precision. First and foremost, we had no direct control over the image acquisition parameters, including illumination intensity, exposure time, or image encoding. With the onboard lighting system being disabled, the delivered laser illumination via the optical fiber was highly dependent on the relative orientation between the fiber output and the target tissue. As a result, illumination uniformity varied considerably across frames, and minor changes in insertion angle or endoscope rotation introduced substantial variations in speckle contrast.
The effective imaging area extracted from the display was a square region of 473$\times$473 pixels, manually cropped from screen captures. Temporal sampling was performed at 60 Hz. However, since we had no access to the native video acquisition or buffering protocols, this framerate does not necessarily reflect the true acquisition frequency of the embedded camera sensor, and it is given here simply as an indication. It may also reflect the display refresh behavior rather than the frame capture process itself. Consequently, the temporal fidelity of the speckle signal is not guaranteed.
Furthermore, the frequency-domain analysis of the recorded speckle sequences revealed the presence of strong spatial aliasing artifacts. These were evidenced by structured peaks in the spatial Fourier transform, consistent with undersampling of high-frequency components and potential resampling artifacts due to unknown digital processing steps within the video pipeline.
Perhaps most critically, we lacked any information regarding the camera’s integration time — a parameter of central importance in dynamic speckle imaging. The integration time has a direct influence on speckle contrast and sensitivity to motion.	Without control or knowledge of this parameter, it is difficult to optimize acquisition settings for either slow or fast biological dynamics. In summary, while the system enabled the acquisition of time-resolved speckle images, its proprietary nature and lack of transparency regarding acquisition parameters pose fundamental limitations to quantitative speckle analysis. These constraints underscore the need for either greater access to raw data or the development of custom acquisition platforms if dynamic speckle is to be pursued
as a clinically viable imaging modality.

\section{Results}

We evaluated the feasibility of dynamic speckle imaging using our adapted endoscopic setup on a superficial vascular lesion (angioma). A sequence of 764 consecutive frames was captured at an estimated cadence of 60 Hz using the previously described screen capture protocol. The region of interest was cropped to a 473 $\times$ 473 pixel window, from which temporal speckle metrics were derived.

\begin{figure}[ht]
    \centering
    \includegraphics[width=0.5\linewidth]{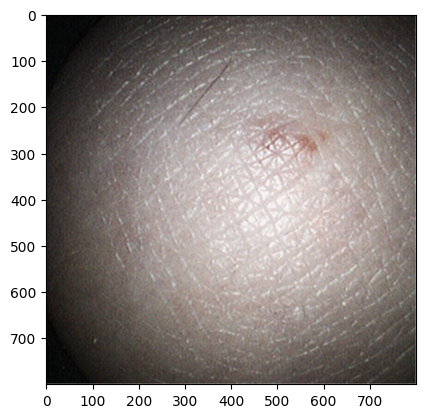}
    \caption{Raw RGB image acquired under standard LED illumination, showing the angioma structure used as a target.}
    \label{fig:rgb}
\end{figure}

Figure~\ref{fig:rgb} shows the RGB image acquired under standard white-light LED illumination from the bronchoscope. This image provides anatomical context of the angioma, but contains no dynamic information.

The core of the analysis consisted in computing the inverse squared speckle contrast, implemented as the squared ratio of temporal mean to standard deviation at each pixel location. This parameter is proportional to established physiological markers such as the Microvascular Activity Index (MAI) \cite{colin2022imaging} or the Blood Flow Index (BFI) \cite{wagner2003reproducibility}, which are commonly used to characterize the level of microcirculatory or cellular activity. These indices are theoretically linked to the inverse of the field decorrelation time and are designed to remain largely independent of absolute illumination intensity. However, a proper normalization by the camera's integration time is typically required to ensure quantitative comparability across conditions \cite{orlik2024standardizing}. In our case, the lack of access to this acquisition parameter precluded such correction, restricting our analysis to a relative, yet physically interpretable, contrast-based measure.

Additionally, illumination across the sequence showed noticeable non-homogeneity due to geometric variation in fiber positioning and possibly tissue bidirectional reflectance distribution function (BRDF).

\begin{figure}[ht]
    \centering
    \includegraphics[width=0.8\linewidth]{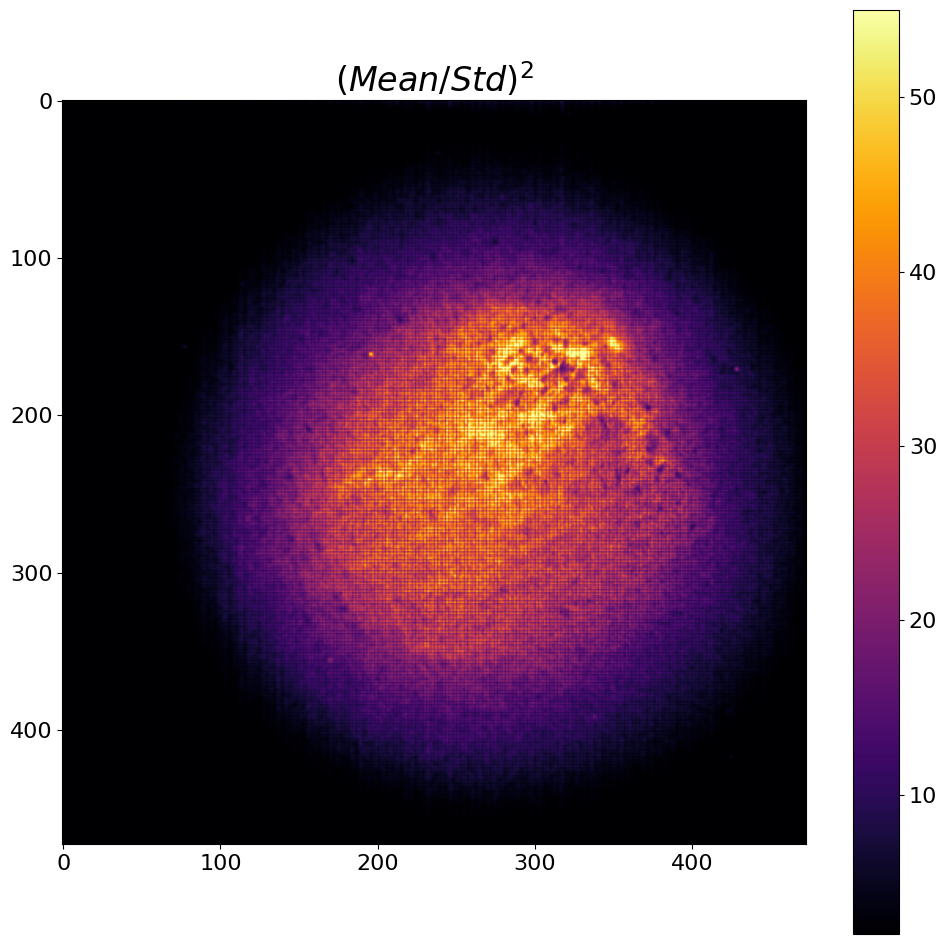}
    \caption{Unprocessed dynamic speckle Activity map (inverse squared speckle contrast) computed over 750 frames. Vascular features are visible, alongside spatial aliasing artifacts.}
    \label{fig:result}
\end{figure}

\begin{figure}[ht]
    \centering
    \includegraphics[width=0.8\linewidth]{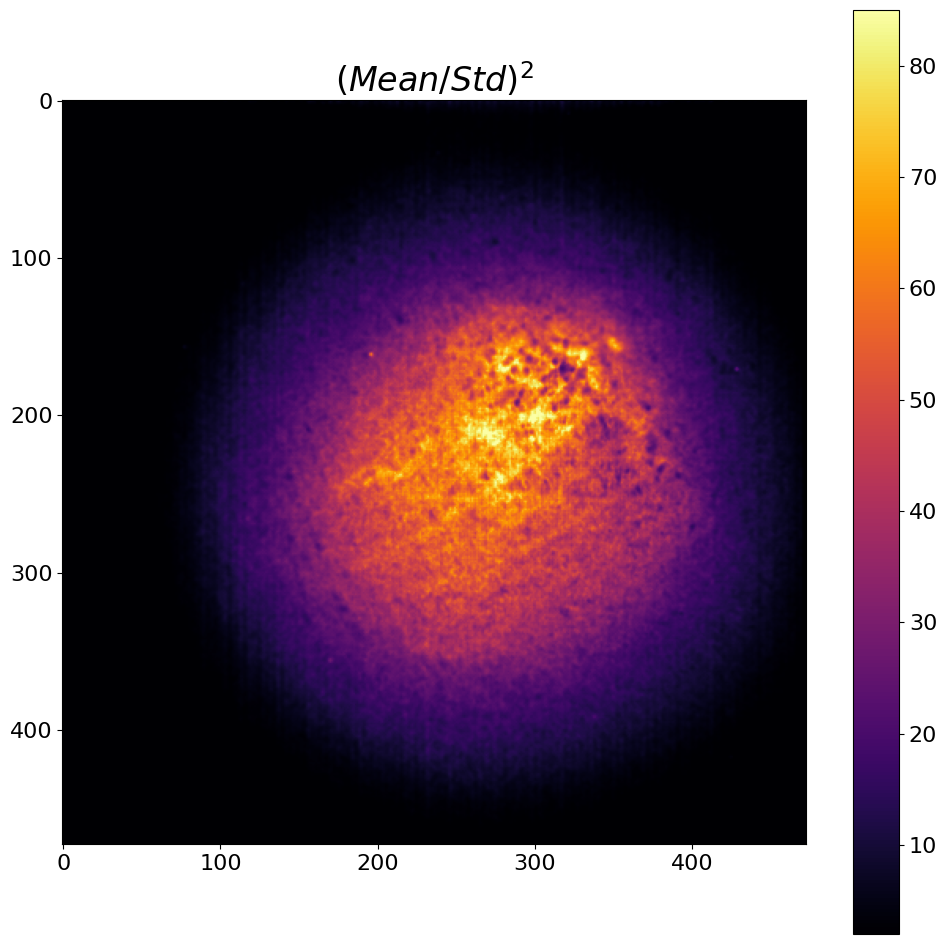}
    \caption{Activity map (inverse squared speckle contrast) after applying a $3\times3$ spatial median filter on each frame before speckle computation. This reduces fine-scale noise and aliasing artifacts.}
    \label{fig:result_f1}
\end{figure}

\begin{figure}[ht]
    \centering
    \includegraphics[width=0.8\linewidth]{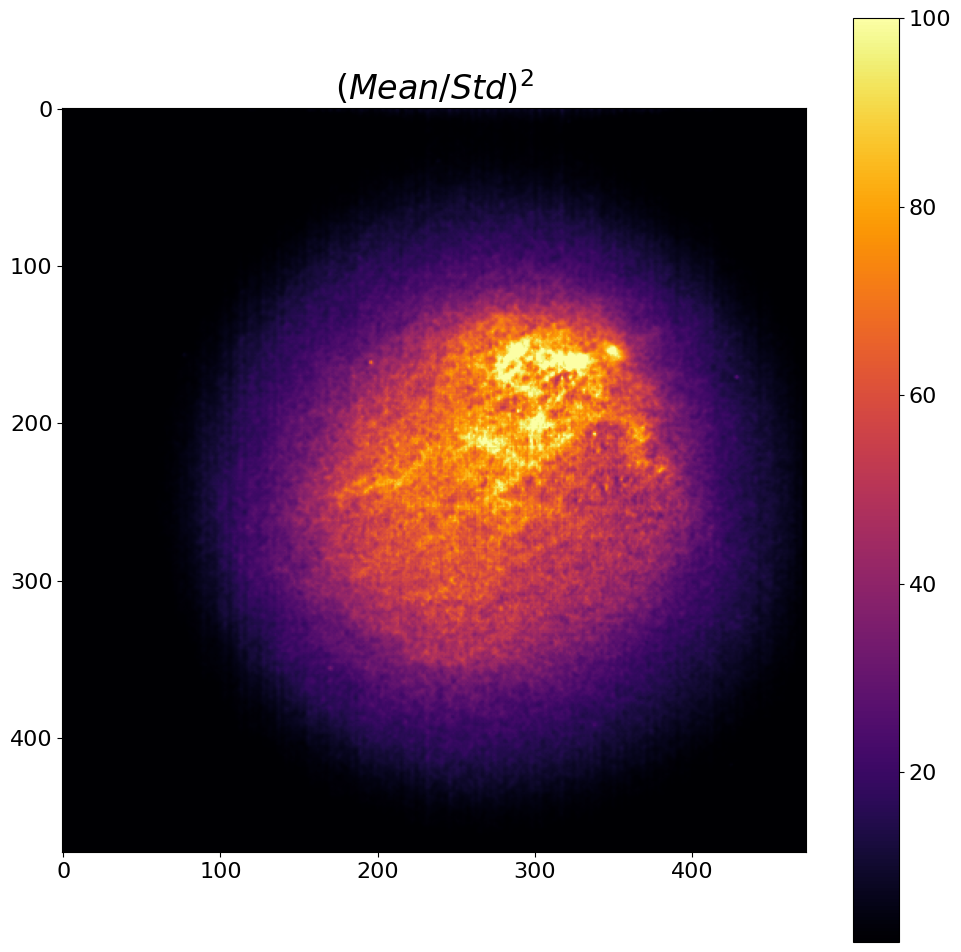}
    \caption{Activity map obtained by computing local inverse squared speckle contrast over successive blocks of 50 frames, followed by temporal averaging. This reduces both the impact of intensity drifts and spatial aliasing.}
    \label{fig:result_f2}
\end{figure}

Despite these limitations, the resulting speckle contrast map (Figure \ref{fig:result}) reveals a noticeable region of elevated activity which corresponds to the location of the angioma (Figure \ref{fig:rgb}) and the related vascular network radiating outward. These features are consistent with microvascular motion and confirm the sensitivity of the method to dynamic processes occurring at least a few hundred microns beneath the skin surface.
However, the visual quality of the MAI-related image shown in Figure \ref{fig:result} is hampered by a parasitic regular grid pattern superimposed on the signal that we attribute to a spatial aliasing artifact present in the raw intensity images that we recorded from EXALT. Application of a Fourier-domain notch filter, which is a common method to attenuate the effects of aliasing \cite{hirano1974design}, proved insufficient to suppress this structure, suggesting its origin is linked to display or compression artifacts before screen capture.
To mitigate these effects, we explored two complementary approaches:
\begin{itemize}
    \item Figure \ref{fig:result_f1}  shows the result afterapplying a 3 × 3 convolution median filter to each frame in the temporal stack before inverse squared speckle contrast computation. This simple and well-known spatial denoising step improved the homogeneity of the output without suppressing the meaningful signal thanks to the reduced size 3x3 of the filter kernel.
\item Figure \ref{fig:result_f2} shows the result of, computing the inverse squared speckle contrast within non-overlapping temporal blocks of 50 frames, and averaged the resulting maps. This approach stabilized the output against slow fluctuations in illumination.
\end{itemize}

These two strategies illustrate the need for a dual-domain approach to improve speckle imaging under constrained acquisition: spatially, to reduce compression- and display-induced aliasing, and temporally, to enhance robustness against non-stationary signal fluctuations.

\section{Discussion}
In the computed activity maps, we consistently observed a pronounced peripheral gradient forming a bluish halo surrounding the central region of interest. This artifact manifests as a lower apparent speckle activity near the image boundaries, despite being theoretically independent of the illumination intensity. Two hypotheses are currently being investigated to explain this phenomenon.

The first hypothesis involves the geometrical configuration of the imaged surface. Due to the inherent curvature and non-planarity of the skin region under investigation, the angle of incidence between the laser illumination and the tissue surface increases progressively from the image center (occupied by the angioma) toward the periphery. As a consequence, the speckle sensitivity to longitudinal (depthwise) motion may decrease in these outer regions. This could be attributed either to the projection of the velocity vector onto the incident light direction being reduced at higher incidence angles or to variations in optical path length between central and peripheral rays, leading to differences in speckle decorrelation dynamics. Further optical modeling and experimental validation are ongoing to quantify this effect.

The second hypothesis involves nonlinearities in the camera response at low light levels. The peripheral regions of the image received less laser illumination due to beam divergence and geometric misalignment. If the camera's response function deviates from linearity under low signal conditions, this could lead to an underestimation of the speckle activity index in those areas. In such a scenario, what appears as a physiological gradient should be an artifact of the sensor's radiometric behavior. To distinguish between these mechanisms and enable quantitative correction, future studies will require a radiometric calibration of the imaging system, as well as controlled acquisitions over planar phantoms and under uniform illumination.

\begin{figure}[ht]
    \centering
    \includegraphics[width=0.6\linewidth]{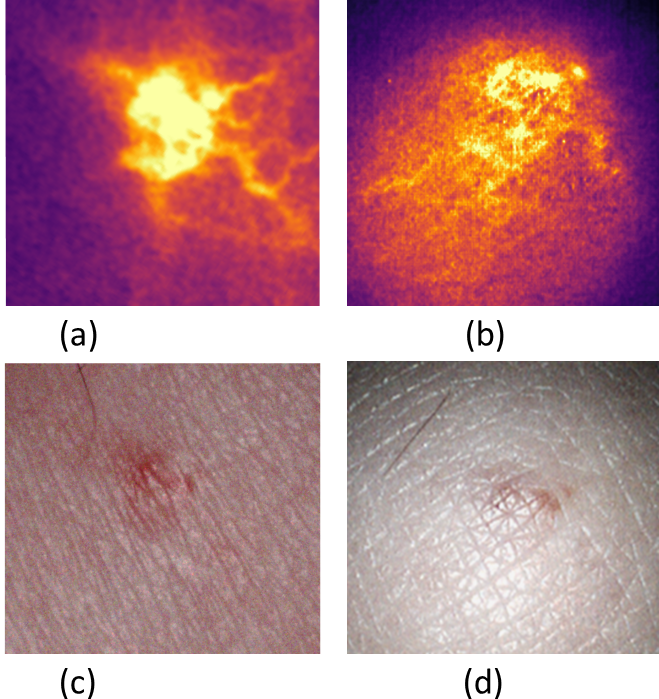}
    \caption{Comparison between conventional vascular imaging and speckle-derived activity mapping using the disposable bronchoscope.
(a) Reference vascular map acquired using a traditional vasculoscope. (b) Speckle activity map computed from temporal fluctuations captured via the disposable bronchoscope. (c) Color image of the angioma obtained under white-light illumination with the vasculoscope. (d) RGB image of the same region acquired via the bronchoscope}
    \label{fig:comparison_f6}
\end{figure}

To qualitatively validate the relevance of the activity patterns revealed by our speckle-based approach, we compared our results with an image acquired using our conventional vascular imaging device (the vasculoscope presented in \cite{colin2022imaging}), under standard illumination conditions. Figure \ref{fig:comparison_f6} shows this side-by-side comparison: on the left, the vascular map recorded with the classical vasculoscope; on the right, the dynamic speckle activity map obtained via the endoscopic signal.

To compensate for the inherent differences in acquisition geometry, most notably a viewing angle mismatch estimated at approximately 45 degrees, we manually rotated the classical speckle image to approximate alignment with the endoscopic field of view. 

Remarkably, several structural correspondences emerge between the two images. Specific vascular branches and central high-activity regions appear spatially coherent, even though direct anatomical correspondence is difficult to establish due to perspective distortion and the absence of rigid registration.

This convergence suggests that the speckle-based index, though acquired under suboptimal and uncontrolled conditions, captures physiologically meaningful vascular information. 

Future work will also focus on motion stabilization, real-time exposure control, and more direct access to unprocessed image streams.

\section{Conclusion}
We present a first proof-of-concept demonstration of dynamic speckle imaging performed through a disposable endoscopic system. This study underscores not only the significant technical limitations associated with using a fully closed and proprietary imaging pipeline, where no access to raw data, acquisition parameters, or exposure control is available, but also the feasibility of extracting physiologically meaningful dynamic information under such constraints.

Remarkably, we obtained these results using an imaging system originally designed for clinical visualization purposes, not for quantitative optical diagnostics. Despite the lack of control over key imaging variables for dynamic speckle imaging, such as illumination stability and homogeneity, integration time precision, compression scheme, and native sensor resolution, dynamic speckle patterns reflecting sub-surface vascular activity were successfully captured.

These findings emphasize the robustness and potential of speckle-based techniques and encourage their adaptation to minimally invasive contexts. Future developments will focus on improving spatial and temporal signal fidelity, mitigating aliasing and instability, and ultimately establishing speckle endoscopy as a viable tool for functional biomedical imaging in constrained clinical environments.

\section*{Acknowledgments}
The authors gratefully acknowledge the support of several individuals and institutions that made this study possible. We thank Dr. Benoît Descantes, Head of the Department of Innovation and Preclinical Research at Hôpitaux Saint Joseph – Marie Lannelongue, for facilitating connections with his clinical and technical teams. Special thanks go to Dr. Adrian Crutu (Interventional Pulmonologist, Hôpital Marie Lannelongue) for initiating the investigation into pulmonary endoscopic imaging and for providing access to the disposable bronchoscope used in this work. We are also indebted to Émilie Bitane, research engineer in Dr. Crutu's team, for her technical assistance and coordination. We extend our appreciation to the surgical team at Hôpital Marie Lannelongue for their collaborative spirit, in particular Ali Akamkam and Pr. Julien Guihaire.

\bibliographystyle{plain}
\bibliography{ref.bib}

\end{document}